\documentclass[onecolumn,pre,aps,superscriptaddress,showpacs]{revtex4}
\begin{document}
\title {scaling law in target-hunting processes}
\author{Shi-Jie Yang}
\address{Department of Physics, Beijing Normal University,
Beijing 100875, China}
\begin{abstract}
We study the hunting process for a target, in which the hunter
tracks the goal by smelling odors it emits. The odor intensity is
supposed to decrease with the distance it diffuses. The Monte
Carlo experiment is carried out on a 2-dimensional square lattice.
Having no idea of the location of the target, the hunter
determines its moves only by random attempts in each direction. By
sorting the searching time in each simulation and introducing a
variable $x$ to reflect the sequence of searching time, we obtain
a curve with a wide plateau, indicating a most probable time of
successfully finding out the target. The simulations reveal a
scaling law for the searching time versus the distance to the
position of the target. The scaling exponent depends on the
sensitivity of the hunter. Our model may be a prototype in
studying such the searching processes as various foods-foraging
behavior of the wild animals.

\end{abstract}
 \maketitle

In the past years, the diffusion-controlled reactions have been
extensively studied through random-walk models. Such applications
range from chemical processes, electronic scavenging and
recombination, to electronic and vibrational energy transfer in
condensed
media\cite{Weiss,Montroll,Klafter,Blumen,Stanley,Szabo,Koza,Bere,Agmon}.
Many works have been devoted to the target annihilation problem,
in which randomly placed targets are annihilated by random
walkers, and its dual of the trapping problem\cite{Weiss2,Jasch}.
Other models treat hindered diffusion problems which involve
random point obstacles \cite{Saxton1}. In these models, the tracer
moves from site to site on a lattice and falls into wells of
various depth at the sites. The tracer does not know the depth of
a well before it enters. Another possibility is a mountain model,
in which all sites are at zero energy and the barriers are on the
bonds joining the sites\cite{Bunde,Saxton3}. Generally,
random-walk models are ideally suited for computer simulations, a
practical way to obtain results, since for the vast majority of
cases no purely analytical method exists.

In this work, we focus on another class of random-walk problems.
We study the so-called target-hunting processes, which frequently
occurs in biological systems, such as a shark searching for foods
by smelling the blood in the ocean, or honeybees flying in the
countryside to locate the
foraging-nectars\cite{Collett,Kareiva,Marsh}, or in metabolic
processes such as cell motions and
chemotaxis\cite{Alt,Keller,Mah,Shenderov,Sherratt}. It can be
viewed as the target-oriented problems, in which the hunters try
to reach the targets by following some kind of behavior rules. In
our model, an active hunter is trying to find out a target which
emits a special kind of odor. The Monte Carlo simulations are
carried out on a 2-dimensional square lattice. Since neither the
distance nor the direction of the target is presumedly to be
known, the searcher should determine its moves by random attempts
in each direction, just like a snake turns its head from side to
side to test the variation of the odor intensity. There are some
chance for the hunter to move in the wrong direction because of
randomness. Hence it is not a traditional biased random-walk.
After sorting each searching process in time sequence, we obtain a
curve with a wide plateau, indicating a most probable time of
successfully finding out the target. By fitting the numerical
results, we find a scaling law for the searching time on the
distance to the position of the target. The scaling exponent is
found to be dependent on the sensitivity of the hunter. We
consider this scaling law scarcely happens in ordinary biased
random walks.

The game rules are as follows:  The hunter at the origin $O$ is
trying to find out a target which emits a special kind of odor.
Since the hunter have no way to know the location of the target,
it randomly moves around its original position to test the
variation of the odor intensity. $z_0$ is the present distance of
the hunter to the target while $z_1$ is the corresponding distance
of the next attempted step. The Monte Carlo steps are implemented
as: if $(z_0/z_1)^\alpha>\zeta$, where the parameter $\alpha$
reflects the sensitivity of the hunter and $\zeta$ is a random
number, then the attempt is accepted. Otherwise it is refused.
This rule implies that the intensity of the signal emitted by the
target is inversely proportional to the distance of the hunter to
the target. Other choices of the relation do not alter the result
qualitatively. By this way, the hunter approaches the goal in a
stochastic style.

Fig.1 displays a typical route of the hunter searching for the
target on a regular lattice. When the hunter is far away from the
goal, the ratio $z_0/z_1$ is close to $1$. Most of the moving
attempts are accepted, even the hunter walks in the wrong
direction. The hunter appears to linger around for quite a while.
Hence the motion of the hunter is nearly a Brownian random walk.
As the goal is nearer, the ratio of $z_0/z_1$ gradually approaches
$0.5$ and the probability of being refused for the hunter moving
in the wrong direction increases. Hence the searching route seems
more straightforward.

Fig.2 shows the searching time for each simulation for a distance
of $z=31.4$ and $\alpha=6$. It is understandable that the
searching time are different for different stochastic processes.
The distribution is not like a white noise. There are large
fluctuations away from the most probable searching time. In Fig.3,
we plot the distribution of searching time for
$z=31.4,65.6,137.1,188.4$, respectively. It is seen that the
distribution is not of the Poissonian form. The curve has a very
long time tail. Instead, a power relation is found for the maximal
value of the distribution with the most probable time,

\begin{equation}
V_{m}\sim t_{p}^{-d_{m}},
\end{equation}
with $d_{m}=1.05$ for $\alpha=6$.

There is another power relation between the distance of the target
and the most probable searching-time,
\begin{equation}
t_p\sim z^{d_p},
\label{dp}
\end{equation}
where $d_p=1.77$ for $\alpha=6$.

In Fig.4 we redistribute the data in Fig.2 by sorting with
increasing time. Fig.4(a) is for various distances $z$ from the
origin, with $z=31.4,65.6,137.1,188.4$ from bottom to top and
$\alpha=6$. The horizonal axis is the sequence of searching time
represented in percentage. A wide plateau is formed in the
intermediate range. Fig.4(b) shows that after proper displacement,
all of the curves collapse into one, implying these curves be
parallel to each other. Hence each curve can be described by a
single function $f(x,\alpha)$ plus a $z$-dependent function $\phi
(z,\alpha)$,
\begin{equation}
\ln t(x,z,\alpha)=f(x,\alpha)+\phi (z,\alpha) \label{ft}
\end{equation}
with $x$ the sequence of searching time represented in percentage.
The function $f(x)$ is of Arabic Ogive-like. It can be checked
that
\begin{equation}
\ln t(\sqrt{z_1\cdot z_2},x)=\frac{1}{2}(\ln t(z_1,x)+\ln
t(z_2,x).
\end{equation}
Formula (\ref{ft}) can be written as
\begin{equation}
\ln t(x,z,\alpha)=f(x,\alpha)+\eta (\alpha)\ln z. \label{fz}
\end{equation}

In figure 5 we studied the dependence of the searching-time with
respect to the sensitivity parameter $\alpha$. Fig.5(a) shows the
curves for $\alpha=4,8,12,16,20$. After properly rescaling the
curves in (a) by times $\ln t$ with a coefficient $\alpha^\beta$,
where $\beta$ is determined below, all of the curves become
parallel. From Fig.5(b), we deduce $\alpha^\beta \ln t=\tilde f
(x)+\tilde \phi (z,\alpha)$. By comparing with eq.(\ref{fz}), one
gets
\begin{equation}
\ln t(x,z,\alpha)=\alpha^{-\beta} f(x)+\eta (\alpha)\ln z.
\end{equation}

The index $\beta$ can be derived by considering the dependence of
the slope $k_2$ of the plateau in figure 5(a) on parameter
$\alpha$. There is a good linear relation between $\ln k_2$ and
$\ln \alpha$, as show in the inset of figure 6,
\begin{equation}
\ln k_2 \sim -\beta \ln \alpha.
\end{equation}
We measured $\beta=0.623$. It is noteworthy that $\beta$ is a
constant independent of sensitivity parameter $\alpha$. It results
from the stochastic process.

Finally, we try to found out the relation between $\eta$ and
$\alpha$. Fig.7(a) depicts the relation of $\ln t$ versus $\ln z$
at $x=0.6$ for various parameter $\alpha$. From figure 7(b),
\begin{equation}
\ln \eta (\alpha)\sim -\delta \alpha
\end{equation}
with $\delta=0.01$.

Combining all the above factors, we consequently obtain a complete
relation of the searching-time with respect to the distance as
well as sensitivity parameter $\alpha$,
\begin{equation}
\ln t(x,z,\alpha)=\alpha^{-\beta}f(x)+c_0 e^{-\delta \alpha}\ln z,
\end{equation}
or
\begin{equation}
t(x,z,\alpha)=e^{f(x)/\alpha^{\beta}}\cdot z^{d}.\label{fr}
\end{equation}

We find that there is a generalized scaling-law between $t$ and
$z$ with exponent
\begin{equation}
d=c_0 e^{-\delta \alpha}.
\end{equation}
From formula (\ref{dp}), $\alpha=6, d=d_p=1.77$. We get
$c_0=1.88$. It shows that the power-law exponent is
$\alpha$-dependent. As $\alpha$ increases from zero to infinity,
the exponent decreases from $1.88$ to zero. In Eq.(\ref{fr}), the
contributions of variable $x$, which sorts the searching time in
each simulation, are completely merged into a prefactor and the
scaling exponent is $x$-independent. It should be noted that it is
a functional relation between the searching time and the distance
in such stochastic processes.

In summery, we introduced a variable $x$ to denote the sequence of
searching time. We plot a curve with a wide plateau, indicating a
most probable time of successfully finding out the goal. In stead
of calculating the mean square root, we introduce a sort parameter
$x$ to figure out an analytical expression. The simulations reveal
a scaling law for the searching time versus the distance to the
position of the target. The scaling exponent is dependent on the
sensitivity of the hunter. We believe that our treatment of the
statistical data may be useful in other cases. The existence of
the scaling law may have implications with the possibility for the
hunter to walk in a wrong direction or stay at the same place for
quite a while. It scarcely happens in an ordinary biased random
walk. We point out that the results are valid not only on the
square lattice, but also for continuous moving (with fixed step
length) in the two-dimensional plane. However, the explicit form
of function $f(x)$ is still lack. It is also desirable to deduce
an analytical express of Eq. (\ref{fr}) from the first principle
of statistics.

We suggest that the scaling law in the hunting process may be an
additional behavior rule in the foods-foraging processes of wild
animals, which has not caught much attention. In turn,
verifications of the law from direct observations by zoologists or
entomologists are also expected. Our target-oriented model may be
a prototype in studying the foods-foraging processes in wildlife
as well as in other searching games.

\centerline {Figure Captions}

Figure 1 A typical hunting route on a square lattice. The start
point is at the origin and the target is at $(25,19)$.

Figure 2 Searching time for 10000 simulations. The original
distance to the target is $z=31.4$ and the sensitivity parameter
$\alpha=6$.

Figure 3 Searching time distribution for distances $z=31.4, 65.6,
137.1, 188.4$, respectively. $\alpha=6$.

Figure 4 Time-sorted curves for 10000 simulations. The horizontal
axis $x$ is the sequence of searching time represented in
percentage and the vertical axis is logarithmic time. $\alpha=6$.
(a) is for various distances $z$ from the origin, with
$z=31.4,65.6,137.1,188.4$ from bottom to top. Evidently, the
curves consist of three parts, and a wide linear region is formed.
These curves is parallel to each other. (b) shows that after
proper displacement, all curves collapse into one.

Figure 5 Time-sorted curves for 10000 simulations for various
parameter values $\alpha$. The horizontal axis $x$ is the sequence
of searching time represented in percentage and the vertical axis
is logarithmic time. The original distance is fixed at $z=31.4$.
(a) is for $\alpha=4,8,12,16,20$ from top to bottom. (b) shows the
rescaled curves of (a) for $\alpha=4,8,12,16,20$ from bottom to
top. These curves are parallel to each other.

Figure 6 The dependence of the slope of the plateau in figure 4 on
parameter $\alpha$. The distance $z=31.4$. Inset: a linear
relation of the logarithmic slope $k_2$ with logarithmic $\alpha$.

Figure 7 (a) A plot of $\ln t$ versus $\ln z$ at $x=0.6$ for
various parameter $\alpha$. (b) A linear relation of $\ln \eta
(\alpha)$ with respect to $\alpha$.

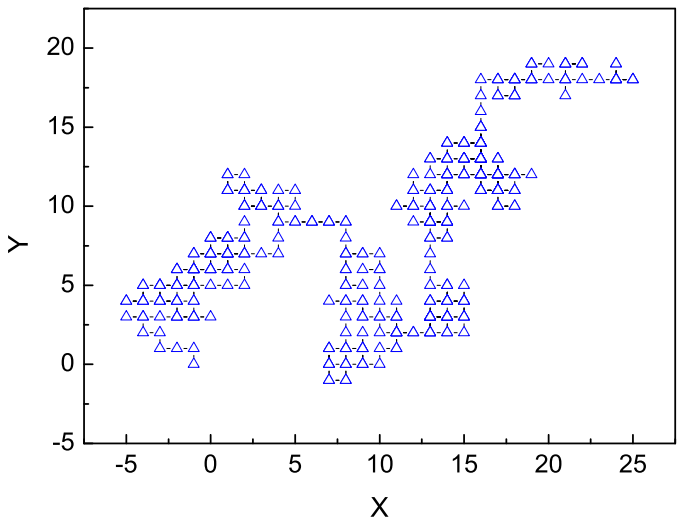
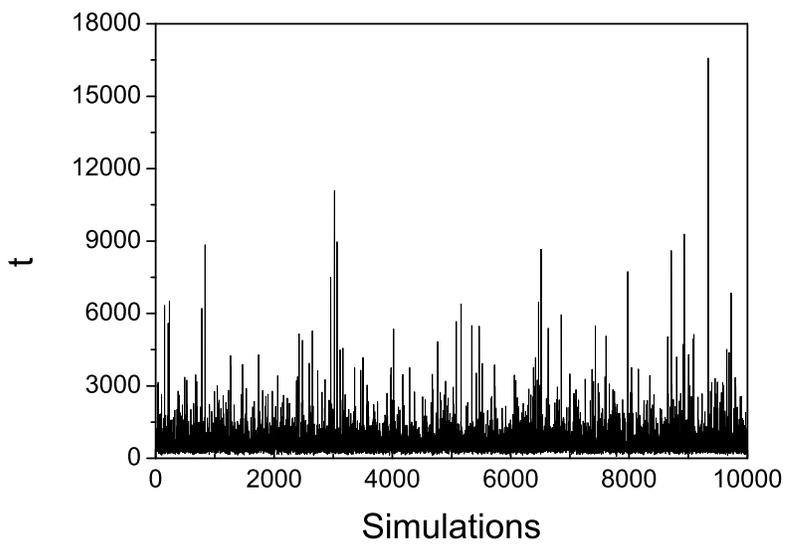
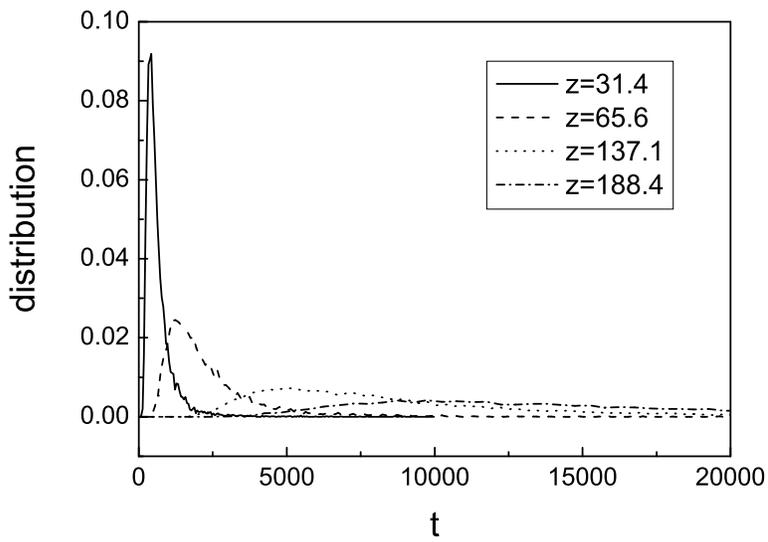
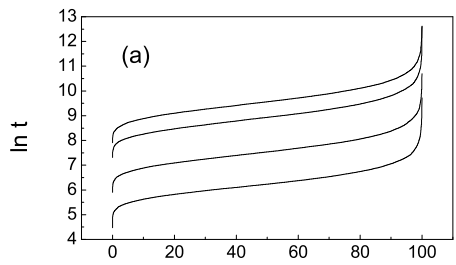
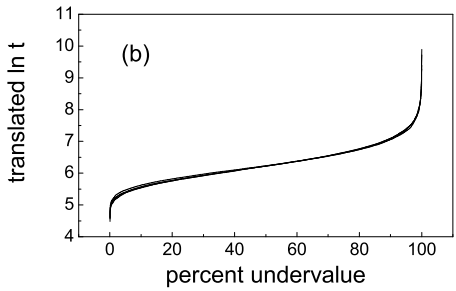
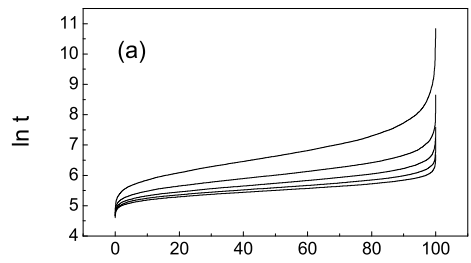
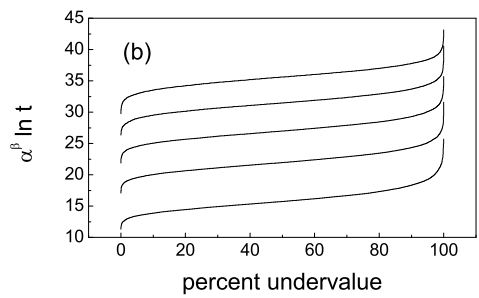
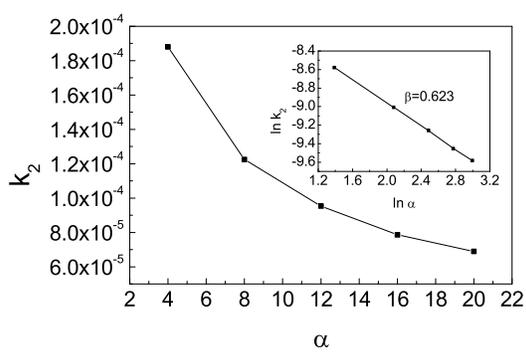
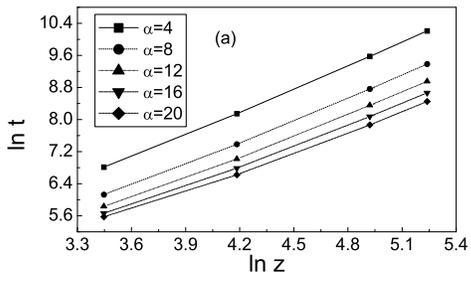
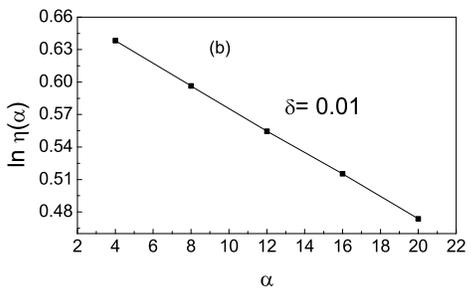
\end{document}